\documentclass[runningheads]{llncs}
\usepackage{graphicx}
\usepackage{algorithm, algorithmic}
\usepackage{subcaption}
\usepackage{multirow}
\usepackage{xcolor}
\begin{document}
\title{CT Image Harmonization for Enhancing Radiomics Studies}
\titlerunning{RadiomicGAN}
%
%
%

\authorrunning{M. Selim et al.}
\author{
Md Selim$^{1,3}$,
Jie Zhang, PhD$^2$,
Baowei Fei, PhD$^{5, 6}$,
Guo-Qiang Zhang, PhD$^7$ ,
Jin Chen, PhD$^{1,3,4}$}

\institute{
Department of Computer Science, University of Kentucky, Lexington, KY \and
Department of Radiology, University of Kentucky, Lexington, KY \and
Institute for Biomedical Informatics,  University of Kentucky, Lexington, KY \and
Department of Internal Medicine,  University of Kentucky, Lexington, KY \and
Department of Bioengineering, University of Texas at Dallas, Richardson, TX \and
Department of Radiology, UT Southwestern Medical Center, Dallas, TX \and
The University of Texas Health Science Center at Houston, Houston, TX \and
}

\maketitle              
\begin{abstract}
While remarkable advances have been made in Computed Tomography (CT), capturing CT images with non-standardized protocols causes low reproducibility regarding radiomic features, forming a barrier on CT image analysis in a large scale. 
RadiomicGAN is developed to effectively mitigate the discrepancy caused by using non-standard reconstruction kernels. 
RadiomicGAN consists of hybrid neural blocks including both pre-trained and trainable layers adopted to learn radiomic feature distributions efficiently. A novel training approach, called Dynamic Window-based Training, has been developed to smoothly transform the pre-trained model to the medical imaging domain. 
Model performance evaluated using 1401 radiomic features show that RadiomicGAN clearly outperforms the state-of-art image standardization models.    

\end{abstract}

\section{Introduction}
\label{sec:intro}\vspace{-0.1in}

As one of the most popular diagnostic image modalities routinely used for assessing anatomical tissue characteristics for disease management~\cite{bushberg2011essential}, computed tomography (CT) provides the flexibility of customizing acquisition and image reconstruction protocols to meet an individual's clinical needs~\cite{midya2018influence}.
However, capturing CT images with non-standardized protocols could result in inconsistent radiomic features in both intra-CT (by changing CT acquisition parameters) and inter-CT (by comparing different scanners with the same acquisition parameters) tests. The low reproducibility regarding radiomic features, such as intensity, shape, and texture, for CT imaging, may forms a barrier to analyzing CT images in a large scale, a.k.a. radiomics~\cite{berenguer2018radiomics,hunter2013high}.

The radiomic feature discrepancy problem can be addressed by either normalizing the radiomic features of all the non-standard images or standardizing CT images and then extracting the radiomic features from the standardized images. The former solution, however, is difficult, since the distributions of the radiomic features are not well defined~\cite{foy2018variation}. For the latter one, image synthesis algorithms have been recently developed aiming to synthesize images with similar feature-based distributions compared to that of the target images while preserving anatomic details~\cite{cohen2012radiosity,ours_aamp,cohen2012radiosity,selim2020stan}. Mathematically, let $x$ be a CT image acquired using a non-standard reconstruction kernel, $y$ be its corresponding standard image, the model aims to compose a synthetic image $y'$ from $x$, such that $y'$ follows the feature distributions of $y$ rather than $x$.
%

Choe et al~\cite{choe2019deep} developed a Convolutional Neural Network (CNN)-based approach for CT image standardization. The model learns the residual representation of the standard images; and then a residual image is combined with a non-standard image to generate a synthesized image. The model, since it trains a CNN from scratch, requires large training data.
Liang et al~\cite{liang2019ganai} proposed a cGAN-based~\cite{pix2pix} CT image standardization model named GANai. A alternative training strategy was developed to effectively learn the data distribution. GANai achieved better performance comparing with cGAN and the traditional histogram matching approach~\cite{gonzalez2012digital}. However, GANai focuses on the relatively easier image patch synthesis problem rather than whole DICOM image synthesis problem.
Selim et al~\cite{selim2020stan} proposed a GAN-based CT image standardization model named STAN-CT. In STAN-CT, a loss function was developed to consider both the latent space loss and the feature space loss. While the former is adopted for the generator to establish a one-to-one mapping from standard images to synthesized images, the latter allows the discriminator to critic the texture features of both standard and synthesized images. Similar to GANai, STAN-CT was applied at image patches and only a few texture features were used as the evaluation criteria.

All these models need to be trained from scratch using a relatively large data which are difficult to obtain. To relax the demand of large data, transfer learning may be adopted. One of the computational challenges is to adopt models pre-trained on natural image domain due to different dynamic ranges. In particular, the pixel intensity of a natural image ranges from 0 to 255 (8-bit), while for the state-of-the-art CT scanners, standard 12-bit depth images are commonly utilized, resulting in a much wider Hounsfield Unit (HU) range that scales from -1,024 to 3,071~\cite{gao2018effects}.

In this paper, we present {\bf RadiomicGAN}, a novel GAN-based deep learning model, for CT image standardization and normalization focused on harmonizing CT images acquired with non-standard reconstruction kernels  as it is one of the most significant factors of feature inconsistency~\cite{choe2019deep}. RadiomicGAN employs a hybrid architecture for image texture feature extraction and embedding.  
%
%
Its encoder consists of multiple consecutive neural blocks including both pre-trained and trainable convolutional layers. 
To address the dynamic pixel range-related problem in transfer learning, RadiomicGAN uses a new training strategy named \textit{Dynamic Window-based Training} (DWT), which allows us to train a model using pixels within a selected range called ``window''. The range of a window can be automatically broaden or shrank based on the pixels where the model suffers most in the previous training iteration, allowing us to fine-tune the trainable layers in RadiomicGAN using the frequently appeared pixels in the window. 

In summary, given its hybrid network structure, RadiomicGAN can effectively learn the radiomic feature distributions from the standard CT images and then harmonize non-standard CT images.
A dynamic window-based training approach is developed to effectively address the pixel range difference problem and thus enable transfer learning in the medical image domain. 

\section{Method} \label{sec_method}\vspace{-0.1in}
%
%

\begin{figure*}[bt!]
\centering
\includegraphics[width=.7\textwidth]{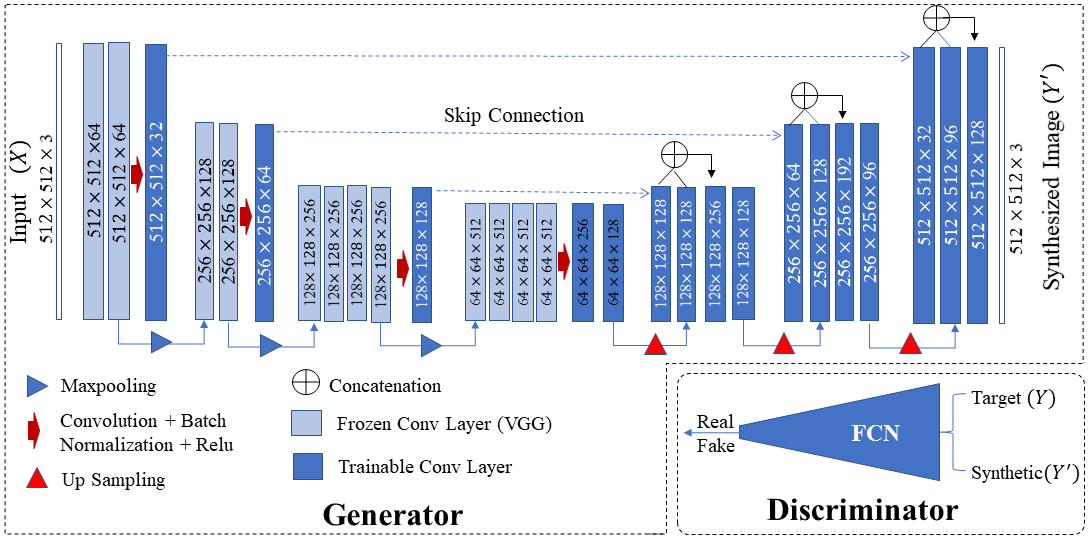}
\caption{\scriptsize \textbf{Architecture of RadiomicGAN.} The generator $G$ is a combination of a U-Net and a pre-trained VGG convolutional layers. The pre-trained frozen layers with trainable layer(s) encode radiomic features effectively, and layers in the decoder reconstruct synthesized images using the encoded features.\vspace{-0.1in}
} \label{fig:model}
\end{figure*}


In RadiomicGAN, by passing a non-standard CT image $x$ through a series of neural blocks, which consists of both pre-trained and trainable convolutional layers, domain-specific texture features of $x$ can be effectively extracted. In the following text, we introduce the hybrid architecture for texture feature extraction and embedding and the dynamic window training, which are the two key components of RadiomicGAN. 

\subsection{Hybrid architecture for efficient texture feature  embedding}\label{sec:GAN}\vspace{-0.1in}
The architecture of RadiomicGAN shown in Figure~\ref{fig:model} consists of a generator $G$ and a discriminator $D$. The discriminator of RadiomicGAN $D$ is a fully convolutional neural network~\cite{pix2pix}.
The encoder of the generator $G$ is constructed as a series of consecutive neural blocks containing both pre-trained and trainable layers inspired by the applications of pre-trained VGG for style transfer~\cite{gatys2015texture,gatys2016image}. Specifically, every neural block includes multiple pre-trained VGG layers (Figure~\ref{fig:model}, light blue), which remain frozen during network training, and a trainable layer (Figure~\ref{fig:model}, dark blue) that works as filter to extract and forward fine-to-coarse texture features to the corresponding decoding layer with a skip connection aiming to preserve the lost features during down-sampling~\cite{u-net}. For example, the first neural block in Figure~\ref{fig:model} includes two 512-by-512 pre-trained layers and a 512-by-512 trainable layer. Both the first and the second layers are constricted by applying the convolution operation with stride=1 on the previous layer. The third layer in the same neural block, however, is trainable and is designed to filter out the redundant features and to propagate domain-specific information onto the corresponding decoding layer. Since we expect to standardize the texture features while keeping the shape features unchanged~\cite{gatys2015texture}, only the first four groups of convolutional layers of a pre-trained VGG-19 network are adopted. 
RadiomicGAN is trained using the adversarial loss with L1 regularization~\cite{pix2pix}. 

\subsection{Dynamic window-based training for efficient information transformation} \label{sec:workflow}
Unlike natural images with pixel values ranging from 0 to 255 (i.e. 8-bit encoding), CT images follow the 16-bit encoding and thus have a much wider dynamic range. 
%
%
%
%
To effectively transfer the learned information from the 8-bit pixel domain to the 12 bit pixel domain, we introduce a new training strategy, called \textit{Dynamic Window-based Training} (DWT). DWT is designed to exposes a deep learning model to the most suitable pixel range in each training iteration, where the lower bound and the upper bound of a window can be automatically updated according to the training results in the previous iterations. 
Compared to the traditional min-max approach that normalizes the whole HU range to $[0,1]$, DWT allows the model to be gradually exposed to the data points with more local intensity details. It enables the utilization of a VGG-19 trained with natural images in ImageNet~\cite{deng2009imagenet} for CT image standardization. 
We introduce two DWT steps, namely fixed growing and dynamic selection, of DWT in the following text. 




\begin{figure} [!bt]
\centering

\includegraphics[width=.8\columnwidth]{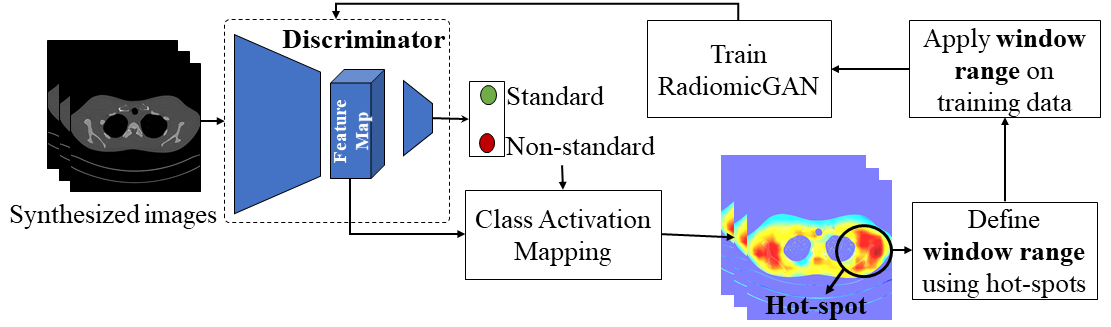}
\caption{\scriptsize \textbf{Dynamic selection framework.} Given a set of CT images synthesized by the generator $G$, the non-trainable discriminator $D$ is used to generate the corresponding class activation mappings (CAMs). A window range is defined using all the hot-spots in the CAMs. RadiomicGAN is then continuously trained using the updated training data.} \label{fig:dynamic-window}\vspace{-0.1in}
\end{figure}

\noindent{\bf Window Training Step 1. Fixed Growing. }
%
%
In the fixed growing approach, we gradually expand the pixel range of a window with a fixed pace. For example, in the first iteration of training, the window only includes HU numbers within a narrow effective pixel range ($[-1024,  -769]$). %
Then, the window size is increased by 256 pixels  (i.e., from $[-1024,  -769]$ to $[-1024, -511]$), allowing for more pixels being considered in the training process. 
Note the window size is increased only if the model accuracy exceeds threshold $th_{acc}$ or reaches maximum training epoch $th_{\eta}$.  
Specifically, for a particular training epoch $n_i$ and a given window $w_i$, the training dataset $DS$ is clipped and normalized based on the min-max value of $w$. With the clipped dataset, the model trains until the model accuracy exceeds threshold $th_{acc}$ or reaches maximum training epoch $th_{\eta}$. Throughout the training process, we repeat the process until the whole HU range is covered. 

  
        


   
        
    	

\noindent{\bf Window Training Step 2. Dynamic Selection. }
Unlike the fixed growing approach, the window range in the dynamic selection approach is determined by using the Class Activation Map (CAM), which identifies the subareas of the input images contributed most for a specific classification task~\cite{zhou2016learning}. Hence, the pace of window expansion or shrinking is not fixed, allowing the model to focus on the areas from where the discriminator of RadiomicGAN $D$ can determine a synthesized image as non-standard. 

The dynamic selection approach is illustrated in Figure~\ref{fig:dynamic-window}. 
Let $\mathcal{P}$ be a randomly selected subset of the training data $DS$ and $\mathcal{P'}$ be its corresponding synthesized images. After each training epoch, we feed $P'$ to the current discriminator $D$ and calculate a set of CAMs $C$ using Grad-CAM ~\cite{selvaraju2017grad}. During the calculation, $D$ remains freeze.  
For each image $p_i\in P'$, the layer before the soft-max layer in $D$ indicates the probability of being a standard and non-standard image, which is used to calculate the CAM $c_i$ of $p_i$. CAM can be visualized as a heat-map where values close to $1$ are critical for determining the image being non-standard. With a user given threshold $th_{fail}$, we can determine the subareas of $p_i$ contributed most for predicting $p_i$ to be a non-standard image ($th_{fail} \in [0, 1]$). 
For image $p_i$, if a pixel's value in the CAM is greater than $th_{fail}$, the corresponding CT image pixel value is added to a pixel intensity list named $W_i$. All the pixels with their frequency equal to $1$ are discarded from $W_i$ considering them as image noise. This process is repeated by randomly selecting subsets of images from the training data. The window range is defined by merging all the pixel intensity lists. 
The window range is recalculated in every epoch and the model is trained based on the current window range.

\section{Experimental Results} \label{sec_experiment}\vspace{-0.1in}
%
%
\subsection{Data, Model, and Evaluation Metric} \vspace{-0.1in}
In total 14,372 CT image slices from lung cancer patient scans and phantom scans were obtained using three different reconstruction kernels (Bl57, Bl64, and Br40) and four different slice thicknesses (0.5, 1, 1.5, 3mm) using the Siemens CT Somatom Force scanner. 
We adopted Bl64 kernel as the standard CT imaging protocol, since it has been widely used in clinical practice for lung cancer diagnosis~\cite{liang2019ganai}. Two testing datasets were prepared for RadiomicGAN performance evaluation. The first testing data were captured using the reconstruction kernel Bl57. The second testing data were captured using the reconstruction kernel Br40. Each test dataset contains 387 image slices and have paired target standard images (Bl64). 
%
%
HU number range was set to between -1024 and 1000 in the standardization process, since most pixel values belong to this range. Random weights were used during the network initialization phase. Maximum training epochs were set to 30 with the learning rate being 0.0001 with momentum 0.5.  

Three state-of-the-art CT image standardization models, i.e. Choe et al.~\cite{choe2019deep}, GANai~\cite{liang2019ganai}, and STAN-CT~\cite{selim2020stan},  were selected for performance comparison. All the models, including RadiomicGAN, were developed based on TensorFlow~\cite{tensorflow2015-whitepaper} and trained using the same training  data. 

Model performance was evaluated systematically at the whole image (DICOM) level and with randomly selected regions of interest (ROIs) in three HU ranges, including $[-800, -300]$, $[-100, 250]$, and $[300, 800]$. 
For each CT image or ROI, in total 1,401 radiomic features were extracted using IBEX~\cite{zhang2015ibex}. 
%

We examined the radiomic features reproducibility performance using Concordance Correlation Coefficient(CCC)~\cite{lawrence1989concordance}. 
CCC represents the correlation between the standard and the non-standard image features in a given features class. CCC ranges from -1 to 1 and is the higher the better. We conclude that a radiomic feature is reproducible if the synthesized image is more than 85\% similar to the corresponding standard image (i.e., $CCC > 0.85$)~\cite{choe2019deep,zhao2016reproducibility}. 
Furthermore, we evaluated image quality using Structural SIMilarity (SSIM)~\cite{wang2004image} and Peak Signal-to-Noise Ratio (PSNR)~\cite{korhonen2012peak}. PSNR is defined as a logged ratio of the peak signal and the mean-square-error between the synthesized and the standard images~\cite{hore2010image} and is the higher the better. SSIM is defined as a co-relation between the synthesized and the standard images with values ranges between -1 to 1, and the value 1 indicates prefect structural similarity.

\subsection{Performance Evaluation}\vspace{-0.1in}
We compared RadiomicGAN with Choe et al, GANai, and STAN-CT using all the three evaluation metrics on both real patient data and phantom data. 
We considered three versions of RadiomicGAN. RadiomicGAN was trained using the proposed window training strategy where both the fixed growing approach and the dynamic selection approach were used consecutively. 
RadiomicGAN$^1$ was a variation trained using the fixed growing approach only, and RadiomicGAN$^2$ was another variation trained using the dynamic selection approach only.
%
%

\begin{table*}[bt!]
 	\footnotesize
 	\centering
 	\caption {\scriptsize CT image standardization model performance comparison for images acquired with Bl57 or Br40 kernels. The values represent the average ($\pm$ SD) number of reproducible radiomic features of the synthesized images. The numbers were rounded to the nearest integer. }
 	\begin{tabular}{| c | c | c | c | c | c | c |}
 		\hline
\textbf{}	&	\multicolumn{3}{c}{Bl57}															&				\multicolumn{3}{|c|}{Br40}		\\ \hline																
\textbf{HU range}	&	[-800, -300] 			&	[-100, 250] 							&	[300, 800]			&				[-800, -300] 			&	[-100, 250] 							&	[300, 800]					\\	\hline
\textbf{Input}	&	854	$\pm$	14	&	589	$\pm$	20					&	457	$\pm$	16	&				448	$\pm$	14	&	303	$\pm$	20					&	246	$\pm$	16			\\	\hline
\textbf{Coe et al.}	&	905	$\pm$	47	&	605	$\pm$	47					&	511	$\pm$	40	&				796	$\pm$	14	&	437	$\pm$	38					&	357	$\pm$	47			\\	\hline
\textbf{GANai} 	&	979	$\pm$	47	&	782	$\pm$	1					&	817	$\pm$	20	&				850	$\pm$	15	&	572	$\pm$	10					&	520	$\pm$	45			\\	\hline
\textbf{STAN-CT}	&	1053	$\pm$	10	&	824	$\pm$	1					&	594	$\pm$	9	&				933	$\pm$	42	&	588	$\pm$	10					&	487	$\pm$	8			\\	\hline
\textbf{RadiomicGAN$^1$}	&	\textbf{1226	$\pm$	47}	&	\textbf{1257	$\pm$	5}					&	\textbf{902	$\pm$	20}	&				1036	$\pm$	25	&	\textbf{1131	$\pm$	29	}				&	1030	$\pm$	15			\\	\hline
\textbf{RadiomicGAN$^2$}	&	1153	$\pm$	33	&	1183	$\pm$	55					&	853	$\pm$	54	&				1087	$\pm$	33	&	1098	$\pm$	44					&	1027	$\pm$	29			\\	\hline
\textbf{RadiomicGAN}	&	1168	$\pm$	36	&	1204	$\pm$	27					&	853	$\pm$	54	&				\textbf{1126	$\pm$	43}	&	1108	$\pm$	25					&	\textbf{1047	$\pm$	36}			\\	\hline

 	\end{tabular} \label{table:comp_feat}
 \end{table*}

Table~\ref{table:comp_feat} indicates the effectiveness of CT image standardization on ROIs with different HU ranges. RadiomicGAN$^1$, RadiomicGAN$^2$, and RadiomicGAN outperformed the models-to-compare on all ROIs. Here, the column named ``Input'' shows the performance of the images before standardization.

In Table~\ref{table:psnr_ssim}, the PSNR scores show that for CT images captured with Br40 or Bl57 kernel, RadiomicGAN achieved the best performance. 

\begin{table*}
\footnotesize
\centering
\caption{\scriptsize PSNR and SSIM scores for images acquired with Bl57 or Br40 kernels.}
	\begin{tabular}{| c | c | c |  c | c | }
	\hline

\textbf{}&	\multicolumn{2}{|c|}{Bl57}								&	\multicolumn{2}{|c|}{Br40}							\\	\hline
\textbf{}	&	PSNR			&	SSIM			&	PSNR			&	SSIM		\\		\hline
\textbf{Coe et al.}	&	27.51	$\pm$	0.90	&	0.9949	$\pm$	0.0038	&	26.11	$\pm$	0.22	&	0.9916	$\pm$	0.0020	\\	\hline
\textbf{GANai} 	&	27.69	$\pm$	0.79	&	0.9951	$\pm$	0.0035	&	26.11	$\pm$	0.22	&	0.9916	$\pm$	0.0021	\\	\hline
\textbf{STAN-CT}	&	27.68	$\pm$	0.74	&	0.9952	$\pm$	0.0022	&	26.08	$\pm$	0.23	&	0.9916	$\pm$	0.0021	\\	\hline
\textbf{RadiomicGAN$^1$}	&	33.12	$\pm$	0.73	&	0.9955	$\pm$	0.0018	&	28.15	$\pm$	0.49	&	0.9918	$\pm$	0.0020	\\	\hline
\textbf{RadiomicGAN$^2$}	&	33.24	$\pm$	0.12	&	0.9955	$\pm$	0.0016	&	30.15	$\pm$	0.16	&	0.9928	$\pm$	0.0015	\\	\hline
\textbf{RadiomicGAN}	&	\textbf{33.51	$\pm$	0.23}	&	\textbf{0.9980	$\pm$	0.0005}	&	\textbf{33.28	$\pm$	0.36}	&	\textbf{0.9947	$\pm$	0.0012}	\\	\hline

	\end{tabular} \label{table:psnr_ssim}\vspace{-0.1in}
\end{table*}

The SSIM scores show that all the models performed well on preserving the structural information. The SSIM scores of RadiomicGAN were slightly higher than the others. 
\begin{figure}
\centering
\includegraphics[width=\columnwidth]{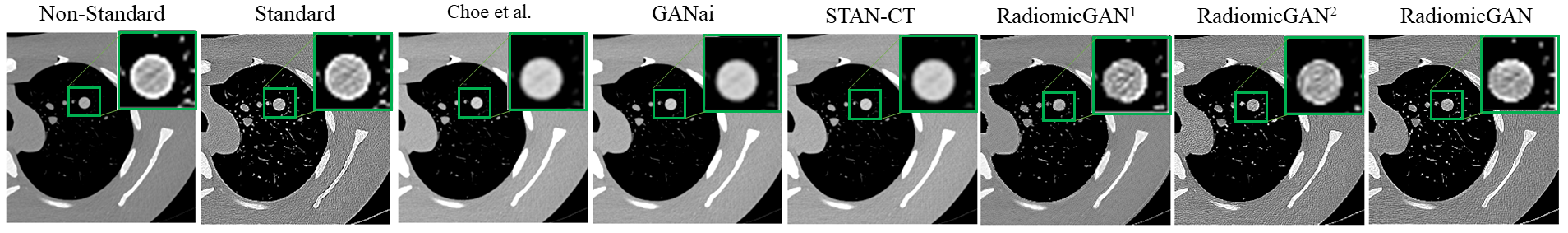}
\caption{\scriptsize CT image standardization case study. An image with a tumor is used as a case study to show the visual quality of all the compared models. The green rectangle highlights a tumor and the display window is [-800, 800] HU.} \label{fig:psnr}\vspace{-0.1in}
\end{figure}

Figure~\ref{fig:psnr} illustrates the performance of all the  models compared using a case study. The non-standard and the standard were phantom CT images acquired with the Br40 and Bl64 kernels respectively. The tumor in the image was highlighted in a green box and was magnified in the right upper corner. The results of all the models were visualized as well. A visual inspection indicates that the tumor image synthesized with RadiomicGAN was the most similar to the standard image. 
The total number of reproducible features computed using CCC was 612 for RadiomicGAN, higher than all the compared methods (Choe et al: 183, GANai: 209, and STAN-CT: 307). The RadiomicGAN$^1$ and RadiomicGAN$^2$ had 487 and 512 reproducible features respectively, which were also better than the compared models. 
The PSNR score of RadiomicGAN was 33.23, clearly higher than the compared methods (Choe et al: 26.08, GANai: 26.14, STAN-CT: 26.07). 

\begin{figure}[bt!]
\centering
\includegraphics[width=.8\columnwidth]{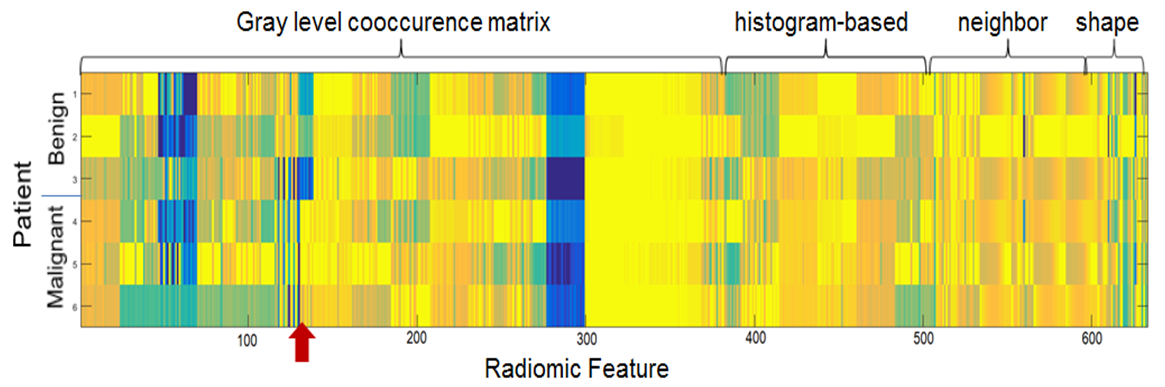} \vspace{-0.1in}
\caption{\scriptsize Heat map representation of image traits for six patients. After image standardization, we found that eight image traits were consistently higher for malignant compared to benign nodules.}\vspace{-0.1in} \label{fig:casestudy}
\end{figure}

While RadiomicGAN has achieved satisfactory performance, it is critical to check whether the image standardization does help cancer diagnosis. Here, we aim to differentiate benign from malignant lung nodules using image features extracted from the standardized CT images. We extracted 630 image features from six CT scans of pediatric sarcoma patients who had undergone resection of a small pulmonary nodule. Three of the resected nodules were histologically confirmed to be malignant (metastatic disease), while the other three were benign. Radiomic evaluation of these nodules identified eight image features consistently high for malignant compared to benign nodules (Figure \ref{fig:casestudy}). In contrast, before image standardization, none of the image features can differentiate malignant and benign nodules. The results suggest that image standardization is critical for CT images acquired with different parameters.  

%

\begin{figure}[bt!]
    \centering
    \subfloat[\scriptsize The window size linearly increases during the fixed growing step and adjusted dynamically in the dynamic selection step. 
]{{\includegraphics[width=.4\columnwidth]{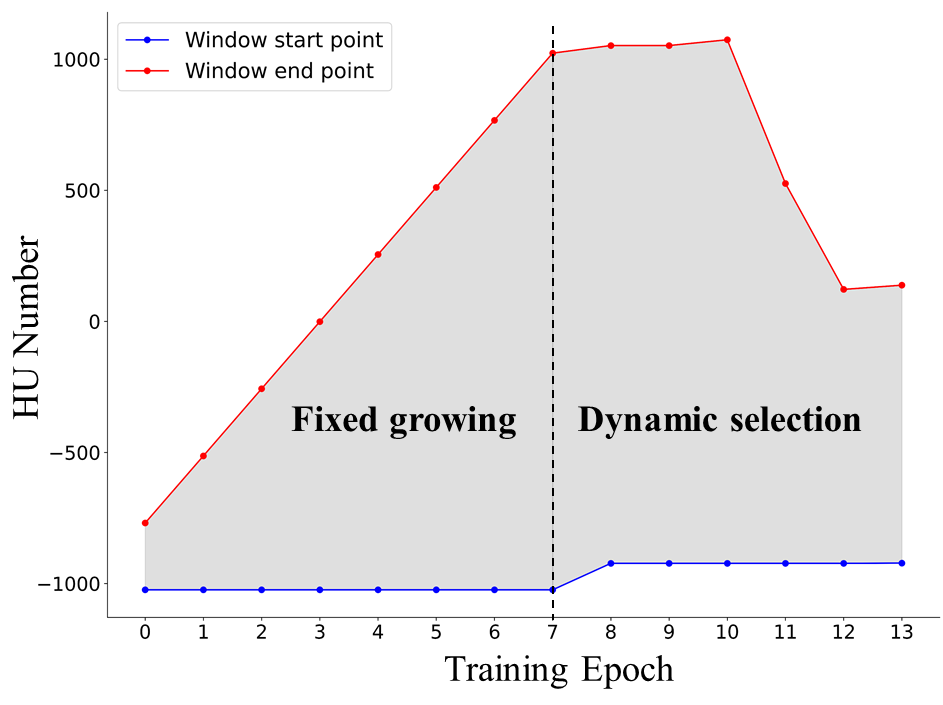} }}
    \qquad
    \subfloat[\scriptsize A sample CAM in the dynamic selection step. The areas corresponding to the active parts (red areas) are the pixels most difficult to synthesize.] {{\includegraphics[width=.45\columnwidth]{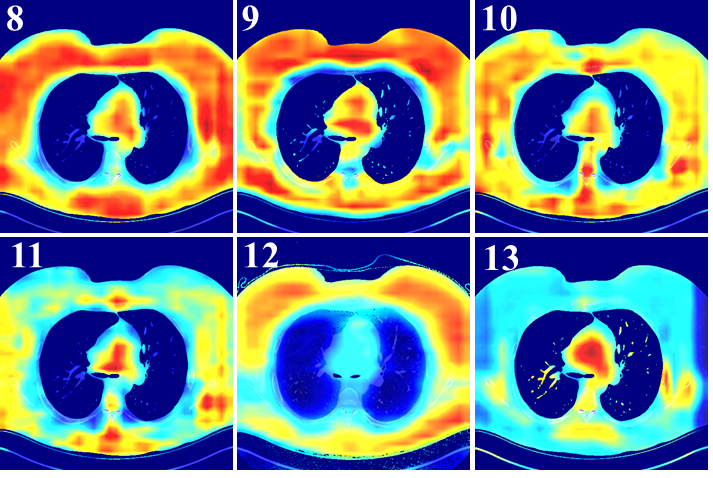} }}
    \caption{\scriptsize Variation of effective pixel range over training epochs.}\vspace{-0.1in}
    \label{fig:grad-cam} \vspace{-0.1in}
\end{figure}

Figure~\ref{fig:grad-cam} illustrate the mechanism and its effect in model training. Figure~\ref{fig:grad-cam} (a) shows the window size variation during training. In the first eight epochs, the model was trained using the fixed growing approach. In epoch 7, the training window range achieved its full dynamic range and shifted the training strategy from the fixed growing approach to the dynamic selection approach. During the dynamic selection approach the window shifted based on the hot-spots of the selected training data. In this phase, the window size reduced dynamically and the training process end at epoch 13 when a reasonable window was reached. 
Figure~\ref{fig:grad-cam} (b) shows a sample image hot-spot during the dynamic selection training phase. The active parts (red areas), which is corresponding to the pixels most difficult to synthesize, reduced quickly, indicating that window training is an effective approach for adopting models trained with ImageNet into the medical imaging domain.  

\section{Conclusion}\vspace{-0.1in}
CT image radiomic feature discrepancy due to the use of non-standard image acquisition kernels creates a gap in large-scale cross-center radiomic studies. RadiomicGAN addresses these challenges by efficiently standardizing and normalizing clinically usable synthetic CT images. The wider dynamic range of CT images has been hindering the adaptation of transfer learning on CT image synthesis. A novel window training strategy is proposed allowing the model to be gradually exposes to the data points with local intensity details, thus significantly improving model performance. In the experiments, we systematically extracted 1,401 radiomic features frequently used in radiomic models. The results show that RadiomicGAN can significantly increase the reproducibility of radiomic features.

\vspace{-0.1in}
\bibliographystyle{splncs04}
\bibliography{main}

\begin{thebibliography}{10}
\providecommand{\url}[1]{\texttt{#1}}
\providecommand{\urlprefix}{URL }
\providecommand{\doi}[1]{https://doi.org/#1}

\bibitem{tensorflow2015-whitepaper}
Abadi, M., Agarwal, A., Barham, P., et~al.: {TensorFlow}: Large-scale machine
  learning on heterogeneous systems (2015), \url{https://www.tensorflow.org/},
  software available from tensorflow.org

\bibitem{berenguer2018radiomics}
Berenguer, R., Pastor-Juan, M.d.R., Canales-V{\'a}zquez, J.,
  Castro-Garc{\'\i}a, M., Villas, M.V., Legorburo, F.M., Sabater, S.: Radiomics
  of ct features may be nonreproducible and redundant: Influence of ct
  acquisition parameters. Radiology p. 172361 (2018)

\bibitem{bushberg2011essential}
Bushberg, J.T., Boone, J.M.: The essential physics of medical imaging.
  Lippincott Williams \& Wilkins (2011)

\bibitem{choe2019deep}
Choe, J., Lee, S.M., Do, K.H., Lee, G., Lee, J.G., Lee, S.M., Seo, J.B.: Deep
  learning--based image conversion of ct reconstruction kernels improves
  radiomics reproducibility for pulmonary nodules or masses. Radiology
  \textbf{292}(2),  365--373 (2019)

\bibitem{cohen2012radiosity}
Cohen, M.F., Wallace, J.R.: Radiosity and realistic image synthesis. Elsevier
  (2012)

\bibitem{deng2009imagenet}
Deng, J., Dong, W., Socher, R., Li, L.J., Li, K., Fei-Fei, L.: Imagenet: A
  large-scale hierarchical image database. In: 2009 IEEE conference on computer
  vision and pattern recognition. pp. 248--255. Ieee (2009)

\bibitem{foy2018variation}
Foy, J.J., Robinson, K.R., Li, H., Giger, M.L., Al-Hallaq, H., Armato, S.G.:
  Variation in algorithm implementation across radiomics software. Journal of
  Medical Imaging  \textbf{5}(4),  044505 (2018)

\bibitem{gao2018effects}
Gao, L., Sun, H., Ni, X., Fang, M., Lin, T.: Effects of 16-bit ct imaging
  scanning conditions for metal implants on radiotherapy dose distribution.
  Oncology letters  \textbf{15}(2),  2373--2379 (2018)

\bibitem{gatys2015texture}
Gatys, L., Ecker, A.S., Bethge, M.: Texture synthesis using convolutional
  neural networks. In: Advances in neural information processing systems. pp.
  262--270 (2015)

\bibitem{gatys2016image}
Gatys, L.A., Ecker, A.S., Bethge, M.: Image style transfer using convolutional
  neural networks. In: Proceedings of the IEEE conference on computer vision
  and pattern recognition. pp. 2414--2423 (2016)

\bibitem{gonzalez2012digital}
Gonzalez, R.C., Woods, R.E.: Digital image processing. Upper Saddle River, NJ:
  Prentice Hall (2012)

\bibitem{hore2010image}
Hore, A., Ziou, D.: Image quality metrics: Psnr vs. ssim. In: 2010 20th
  international conference on pattern recognition. pp. 2366--2369. IEEE (2010)

\bibitem{hunter2013high}
Hunter, L.A., Krafft, S., Stingo, F., Choi, H., Martel, M.K., Kry, S.F., Court,
  L.E.: High quality machine-robust image features: Identification in nonsmall
  cell lung cancer computed tomography images. Medical physics  \textbf{40}(12)
  (2013)

\bibitem{pix2pix}
Isola, P., Zhu, J.Y., Zhou, T., Efros, A.A.: Image-to-image translation with
  conditional adversarial networks. In: Computer Vision and Pattern Recognition
  (CVPR) (2017)

\bibitem{korhonen2012peak}
Korhonen, J., You, J.: Peak signal-to-noise ratio revisited: Is simple
  beautiful? In: 2012 Fourth International Workshop on Quality of Multimedia
  Experience. pp. 37--38. IEEE (2012)

\bibitem{lawrence1989concordance}
Lawrence, I., Lin, K.: A concordance correlation coefficient to evaluate
  reproducibility. Biometrics pp. 255--268 (1989)

\bibitem{ours_aamp}
Liang, G., Zhang, J., Brooks, M., Howard, J., Chen, J.: radiomic features of
  lung cancer and their dependency on ct image acquisition parameters. Medical
  Physics  \textbf{44}(6), ~3024 (2017)

\bibitem{liang2019ganai}
Liang, G., Fouladvand, S., Zhang, J., Brooks, M.A., Jacobs, N., Chen, J.:
  Ganai: Standardizing ct images using generative adversarial network with
  alternative improvement. In: 2019 IEEE International Conference on Healthcare
  Informatics (ICHI). pp. 1--11. IEEE (2019)

\bibitem{midya2018influence}
Midya, A., Chakraborty, J., G{\"o}nen, M., Do, R.K., Simpson, A.L.: Influence
  of ct acquisition and reconstruction parameters on radiomic feature
  reproducibility. Journal of Medical Imaging  \textbf{5}(1),  011020 (2018)

\bibitem{u-net}
Ronneberger, O., Fischer, P., Brox, T.: U-net: Convolutional networks for
  biomedical image segmentation. In: Medical Image Computing and
  Computer-Assisted Intervention. pp. 234--241. Springer (2015)

\bibitem{selim2020stan}
Selim, M., Zhang, J., Fei, B., Zhang, G.Q., Chen, J.: Stan-ct: Standardizing ct
  image using generative adversarial network. In: AMIA Annual Symposium
  Proceedings. vol.~2020. American Medical Informatics Association (2020)

\bibitem{selvaraju2017grad}
Selvaraju, R.R., Cogswell, M., Das, A., Vedantam, R., Parikh, D., Batra, D.:
  Grad-cam: Visual explanations from deep networks via gradient-based
  localization. In: Proceedings of the IEEE international conference on
  computer vision. pp. 618--626 (2017)

\bibitem{wang2004image}
Wang, Z., Bovik, A.C., Sheikh, H.R., Simoncelli, E.P.: Image quality
  assessment: from error visibility to structural similarity. IEEE transactions
  on image processing  \textbf{13}(4),  600--612 (2004)

\bibitem{zhang2015ibex}
Zhang, L., Fried, D.V., Fave, X.J., Hunter, L.A., Yang, J., Court, L.E.: Ibex:
  an open infrastructure software platform to facilitate collaborative work in
  radiomics. Medical physics  \textbf{42}(3),  1341--1353 (2015)

\bibitem{zhao2016reproducibility}
Zhao, B., Tan, Y., Tsai, W.Y., Qi, J., Xie, C., Lu, L., Schwartz, L.H.:
  Reproducibility of radiomics for deciphering tumor phenotype with imaging.
  Scientific reports  \textbf{6}(1), ~1--7 (2016)

\bibitem{zhou2016learning}
Zhou, B., Khosla, A., Lapedriza, A., Oliva, A., Torralba, A.: Learning deep
  features for discriminative localization. In: Proceedings of the IEEE
  conference on computer vision and pattern recognition. pp. 2921--2929 (2016)

\end{thebibliography}
%




\end{document}